\documentclass[aps, twocolumn, 10pt, showpacs, longbibliography, superscriptaddress]{revtex4-1}
\usepackage{graphicx}% Include figure files
\usepackage{lipsum}
\usepackage{dcolumn}% Align table columns on decimal point
\usepackage{bm}% bold math
\usepackage{color}
\usepackage{amsmath, amsthm, amssymb}
\usepackage[modulo]{lineno}%[switch]{lineno} 
\usepackage[usenames,dvipsnames]{xcolor}
\usepackage[normalem]{ulem}
\newcommand{\ba}{\begin{eqnarray}}
\newcommand{\ea}{\end{eqnarray}}

\usepackage[colorlinks=true, urlcolor=violet, linkcolor=blue, citecolor=red, hyperindex=true, linktocpage=true]{hyperref}

\usepackage{float}
\floatstyle{boxed}
\begin{document}
\title{Josephson-junction infrared single-photon detector}
%% Notice placement of commas and superscripts and use of &
%% in the author list

\author{Evan D. Walsh}
\affiliation{Department of Electrical Engineering and Computer Science, Massachusetts Institute of Technology, Cambridge, MA 02139}
\affiliation{School of Engineering and Applied Sciences, Harvard University, Cambridge, MA 02138}
\author{Woochan Jung}
\affiliation{Department of Physics, Pohang University of Science and Technology, Pohang 790-784, Republic of Korea}
\author{Gil-Ho Lee}
\affiliation{Department of Physics, Pohang University of Science and Technology, Pohang 790-784, Republic of Korea}
\affiliation{Department of Physics, Harvard University, Cambridge, MA 02138}
\author{Dmitri K. Efetov}
\affiliation{ICFO-Institut de Ci\`{e}ncies Fot\`{o}niques, The Barcelona Institute of Science and Technology, 08860 Castelldefels (Barcelona), Spain}
\author{Bae-Ian Wu}
\affiliation{Air Force Research Laboratory, Wright-Patterson AFB, OH 45433}
\author{K.-F. Huang}
\affiliation{Department of Physics, Harvard University, Cambridge, MA 02138}
\author{Thomas A. Ohki}
\affiliation{Raytheon BBN Technologies, Quantum Engineering and Computing Group, Cambridge, Massachusetts 02138, USA}
\author{Takashi Taniguchi}
\affiliation{International Center for Materials Nanoarchitectonics, National Institute for Materials Science, Tsukuba, Japan}
\author{Kenji Watanabe}
\affiliation{Research Center for Functional Materials, National Institute for Materials Science, Tsukuba, Japan}
\author{Philip Kim}
\affiliation{Department of Physics, Harvard University, Cambridge, MA 02138}
\author{Dirk Englund}
\affiliation{Department of Electrical Engineering and Computer Science, Massachusetts Institute of Technology, Cambridge, MA 02139}
\author{Kin Chung Fong}
\email{fongkc@gmail.com}
\affiliation{Raytheon BBN Technologies, Quantum Engineering and Computing Group, Cambridge, Massachusetts 02138, USA}

\date{\today}
\begin{abstract}
Josephson junctions (JJs) are ubiquitous superconducting devices, enabling high sensitivity magnetometers and voltage amplifiers, as well as forming the basis of high performance cryogenic computer and superconducting quantum computers. While JJ performance can be degraded by quasiparticles (QPs) formed from broken Cooper pairs, this phenomenon also opens opportunities to sensitively detect electromagnetic radiation. Here we demonstrate single near-infrared photon detection by coupling photons to the localized surface plasmons of a graphene-based JJ. Using the photon-induced switching statistics of the current-biased JJ, we reveal the critical role of QPs generated by the absorbed photon in the detection mechanism. The photon-sensitive JJ will enable a high-speed, low-power optical interconnect for future JJ-based computing architectures.
\end{abstract}
%\pacs{65.80.Ck, 68.65.-k, and 07.57.Kp}
\maketitle 
%\linenumbers
%\setstretch{2}
%\doublespacing
Exploiting its macroscopic quantum behavior, the Josephson junction (JJ) is arguably the most important superconducting device with its wide array of applications: high sensitivity magnetometers \cite{CLARKE:1974gs}, quantum noise limited microwave parametric amplifiers \cite{CastellanosBeltran:2008ke,Yamamoto:2008cra}, rapid single flux quantum in high performance cryogenic computer \cite{Chen:1999fj,Hashimoto:2009db}, and qubits in superconducting quantum computers \cite{Martinis:2003bq,Riste:2013il,Wang:2014dv,Serniak:2018dp,Vepsalainen:2020ij}. While JJ performance can be degraded by the quasiparticles generated from breaking of Cooper pairs with radiation as observed in qubit relaxation \cite{Vepsalainen:2020ij}, the phenomenon also opens opportunities to high sensitivity photodetection. JJ photodetection has been pursued since the earliest realization of JJs \cite{Giaever:1968is,CLARKE:1974gs}. However, despite numerous research efforts to utilize mechanisms such as non-linearity in the AC Josephson effect, non-equilibrium superconductivity, photo-induced carriers, and the bolometric effect \cite{McGrath:1998jj,Schapers:1999gda,Stella:2008dh,Wang:2015by,Tsumura:2016ksc}, single-photon detection by JJs remains elusive. In fact, a superconducting tunnel junction (STJ) \cite{Peacock:1996io,Segall:2000cd} can only be single-photon sensitive by deliberately suppressing the Josephson coupling with an external magnetic field. Here, we demonstrate near-infrared (NIR) single-photon detection by current-biased JJs. By coupling photons to a lateral proximity JJ using localized surface plasmons \cite{Engheta:2007ca}, we measure the single-photon induced JJ switching as a function of current bias, temperature, photon rate, and polarization. Our analysis indicates the JJ switching is caused by the QPs produced from the absorption of a single photon in the superconductor, clearly distinguished from bolometric effects \cite{Giazotto:2008gd,Oelsner:2013iw,Tsumura:2016ksc,Lee:2020ci,Kokkoniemi:2020dk}. Compared to other superconductor-based single-photon detectors (SPD) \cite{Goltsman:2001eaa,Korzh:2020jo,Sadleir:2010bf,Karasik:2012uv,Echternach:2018iw}, our single-photon sensitive JJ is more readily integratable into future JJ-based computing architectures as a high-speed, low-power optical interconnect. Our result also provides insights into protecting qubits from relaxation induced by photon-generated QPs.

\begin{figure*}
\includegraphics[width=5in]{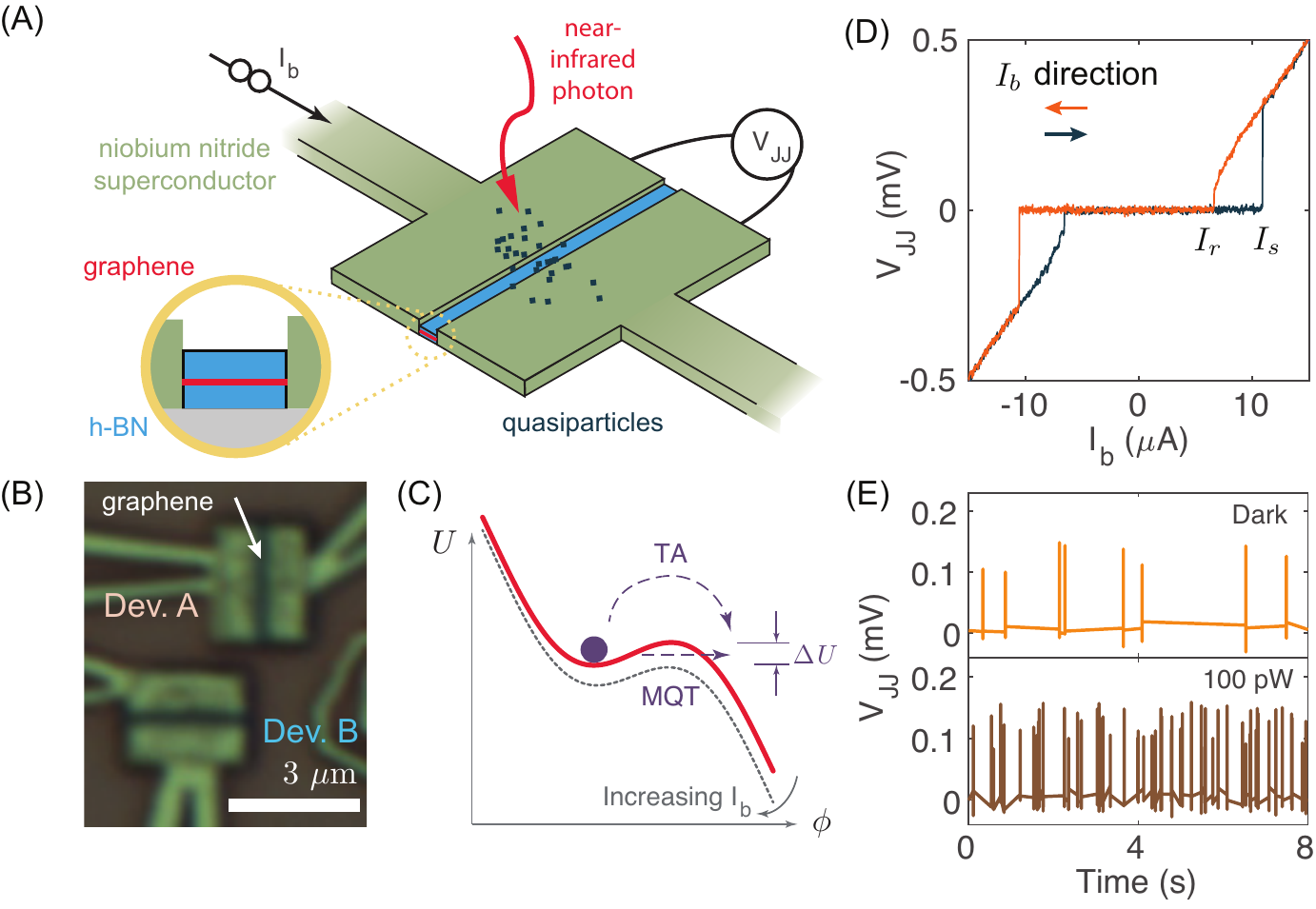}
\caption{\textbf{Current-biased Josephson junction (JJ) as single-photon detector (A)} Illustration of the photon-JJ interaction. Quasiparticles (QPs) are generated as incident photons break Cooper pairs and trigger the JJ to switch. \textbf{(B)} Optical image of two graphene-based JJs. Each JJ is made of a hBN-encapsulated graphene strip of length 160 nm by width 2.8 $\mu$m contacted on each end by NbN pads of the same width and 1 $\mu$m length.  \textbf{(C)} Current-biased JJ can be described as a macroscopic quantum phase particle subjected to a tilted-washboard potential in RCSJ model. The phase particle can be driven into motion, representing the normal resistive state, from the stationary state, representing the supercurrent state, by thermal activation (TA) or macroscopic quantum tunneling (MQT). Additional noise from QP diffusion can enhance the state transition and thus the JJ switching probability. \textbf{(D)} Typical IV curve displaying the switching of JJ between supercurrent and normal resistive state. \textbf{(E)} Switching events in the absence (top) and presence (bottom) of 100 pW of 1550-nm light at $I_b = 10.90 \mu$A.}
\end{figure*}
Schematically shown in Fig. 1A, we investigate the photon-JJ interaction by illuminating the JJ (Fig. 1B) with a 1550 nm NIR laser, brought via a single-mode optical fiber into the dilution fridge at 27 mK. The proximity JJ is fabricated \cite{Walsh:2017kk} by depositing superconductor, i.e. 5-nm niobium (Nb) and 50-nm niobium nitride (NbN) with 5-nm titanium as an adhesive layer, on the sides of graphene which is encapsulated between two atomically flat and insulating hexagonal boron nitride layers. In contrast to trilayer JJs, lateral JJs expose and couple directly to the NIR photons. Upon being absorbed into the superconductor, a single-NIR photon will break Cooper pairs and generate QPs, which then become a noise source to switch the current-biased JJ. The probability of JJ switching can be described by the resistively and capacitively shunted junction (RCSJ) model \cite{Tinkham:book}, in which a macroscopic quantum phase particle with a phase difference, $\phi$, between the two superconducting electrodes is subject to a washboard potential (Fig. 1C). When the phase particle is trapped initially in a local minima, i.e. $d\phi/dt = 0$, the voltage drop across the JJ is zero (Fig. 1D). The bias current $I_b$ running through the JJ tilts the washboard potential and the phase particle could escape from the metastable minimum. When it escapes, either by thermal activation (TA) \cite{Martinis:1987ua} over or macroscopic quantum tunneling (MQT) \cite{Devoret:1985jx} through the barrier ($\Delta U$), the voltage drop across the JJ becomes finite and the JJ switches to the normal resistive state at a switching current $I_s$. The phase particle can be retrapped at a retrapping current $I_r$ by ramping down $I_b$. The hysteretic behavior, i.e. $I_s>I_r$, frequently observed in graphene-based JJs due to self-Joule heating \cite{Courtois:2008iv,Borzenets:2016gw,Lee:2020ci}, is useful to our investigation. When the JJ latches into the resistive state after switching, we can register a count, reset the bias current, and over time, measure the switching statistics under different light intensities and conditions  \cite{Walsh:2017kk}. As Fig. 1E shows, there are considerably more switching events with even just 100-pW of illumination. We studied seven different JJs [Supplementary Information (SI)] that detect single-photons, but we will present results mostly from one device (Device A) and compare it with the rest as controls to understand the JJ interaction with single NIR photons.

\begin{figure}
\includegraphics[width=0.7\columnwidth]{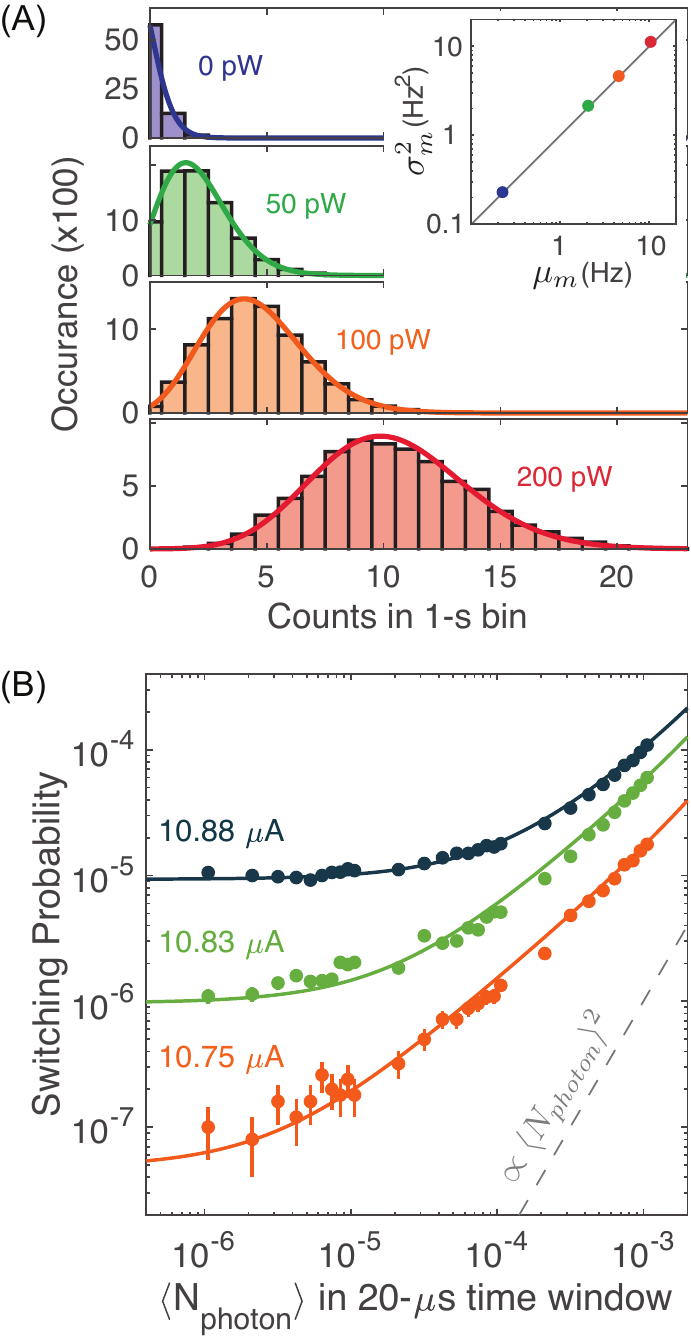}
\caption{\textbf{Switching of Josephson junction (JJ) by single NIR photon. (A)} The observed switching events under illumination follow Poisson statistics showing the events are uncorrelated. Shown from top to bottom are the histograms of switching counts in 1-s bins for 0, 50, 100, and 200 pW of illumination. Inset: The statistical variance of counts $\sigma^2_m$, i.e. the photon shot noise, equals to its mean $\mu_m$. Solid line shows the theory without fitting parameter. \textbf{(B)} The switching probability as a function of average absorbed photon number $\langle N_{photon}\rangle$ ($\ll 1$) in 1-ms time window for three bias currents. Switching probability is linearly proportional to $\langle N_{photon}\rangle$ with an offset due to dark count. The solid line shows the fitting weighted with the inverse of standard deviation (error bars). The linearity proves the switching of the JJ is induced by single photons.}
\end{figure}
The seemingly random switching with light illumination actually follows the statistics of photon shot noise. After taking 10$^4$-s time traces at various powers, we produce histograms of switching events using 1-s bins in Fig. 2A. A Poisson distribution (solid lines) traces the experimental data closely (bars) as expected for uncorrelated switching events. Without any fitting, the data in the inset also demonstrates the property of Poisson statistics that the variance of the count is equal to the mean. Furthermore, we find the switching probability in a 1-ms long time window, accounting for an offset due to the false positive (dark) count, depends linearly on the average absorbed photon number, $\langle N_{photon}\rangle$. $\langle N_{photon}\rangle$ is calculated using the rate of absorbed photons by the JJ, $\mathcal{R}_{photon}$, i.e. 53 Hz at 100 pW laser power, estimated from the polarization measurement discussed later. Fig. 2B shows this property over a large range of $\langle N_{photon}\rangle$ and at several values of $I_b$. Higher $I_b$ produces higher switching probability because of higher intrinsic quantum efficiency and dark count \cite{Walsh:2017kk}. Since the Poisson probability for measuring $m$ photons in a detection time window reduces to $\simeq\langle N_{photon}\rangle^m/m!$ for $\langle N_{photon}\rangle\ll 1$, this linearity proves that our JJ detects single NIR photons from a dim coherent source \cite{Goltsman:2001eaa}. 

\begin{figure}
\includegraphics[width=0.7\columnwidth]{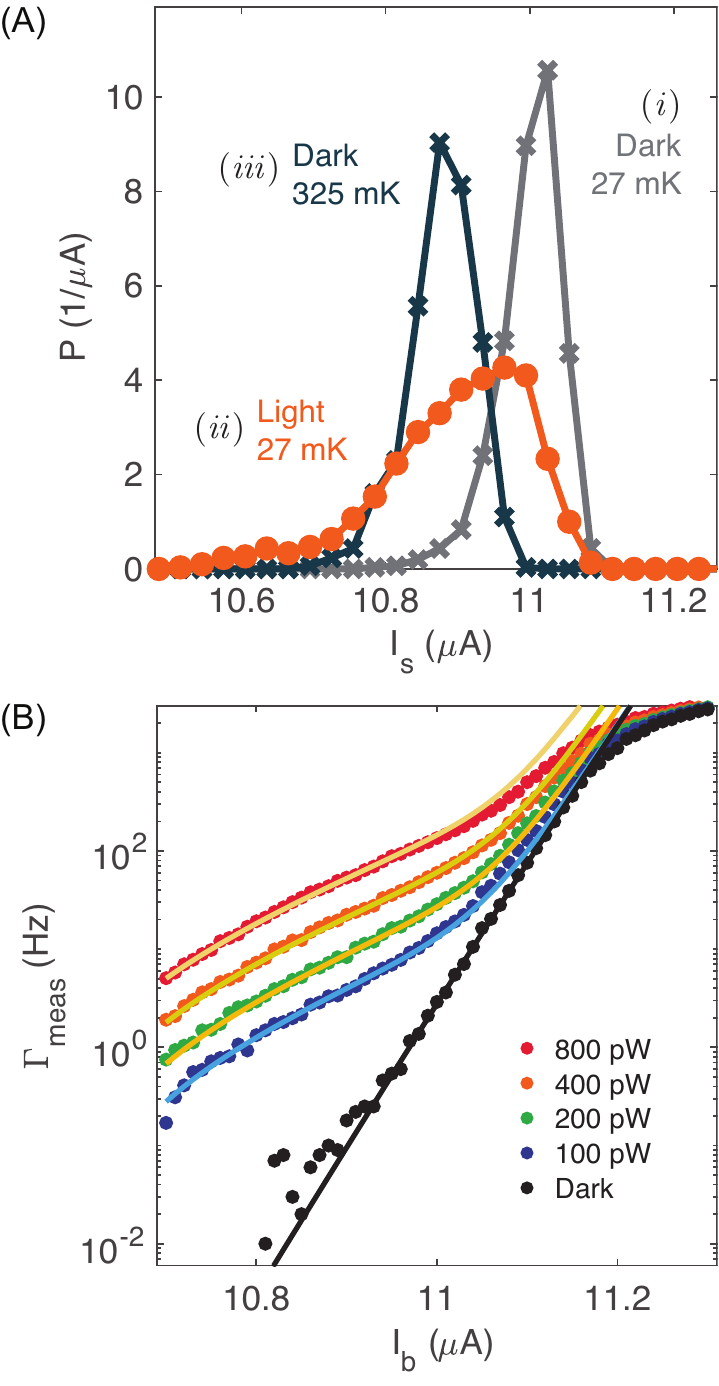}
\caption{\textbf{Switching mechanism. (A)} The distribution of switching current $I_s$ by ramping up $I_b$ at 1 $\mu$A/s under three conditions: (\textit{i}) at 27 mK in the absence and (\textit{ii}) presence (yellow) of 100 pW of 1550-nm light, and (\textit{iii}) at 325 mK in the absence of light. Although the 27-mK, 100-pW distribution has the same averaged $I_s$ as the 325-mK distribution in the dark, the shapes of the two distributions are very different, ruling out steady heating by laser power as the mechanism of the observed increased JJ switching rate. \textbf{(B)} The switching rate as a function of bias current without light and with various incident laser powers. The points are measured values while the black line is the best fit of the MQT process to $\Gamma_{meas}$ without light and the color lines are modeled results from photon-enhanced $\Gamma^{light}_{esc}$. The linearity of the data on a semilog plot indicates the enhancement is due to a momentary elevation of the activation energy by the photon over the potential barrier $\Delta U$ in the reaction-rate theory.}
\end{figure}
It is imperative to distinguish JJ switching induced by single photons from steady heating by laser power. Therefore, we compare the $I_s$ distributions by ramping up $I_b$ for 1000 times (Fig. 3A) under three conditions: (\textit{i}) at 27 mK without light, (\textit{ii}) at 27 mK with light, and (\textit{iii}) at 325 mK without light. The $I_s$ distribution reflects the stochasticity of the phase particle escape in the tilted-washboard potential and measures distinctively the phase particle escape rate without light, $\Gamma_{esc}^{(dark)}$ \cite{Fulton:1974uf}. While both the average values $\langle I_s\rangle_{(ii)}$ and $\langle I_s\rangle_{(iii)}$ are suppressed equally, by photon illumination and operating at a higher temperature, respectively, they do not have the same shape distribution. This result contrasts sharply to the previous experiments in which JJs under photon illumination had $I_s$ distributions matching those of JJs with elevated temperature \cite{Wang:2015by,Lee:2020ci}. The difference of $I_s$ distributions in Fig. 3A excludes the steady heating of the JJ by laser power as the cause of increased JJ switching.

To study the switching mechanism, we compare the measured switching rate, $\Gamma_{meas}$, versus $I_b$ with and without illumination (Fig. 3B). Without light, $\Gamma^{(dark)}_{meas}$ directly measures $\Gamma^{(dark)}_{esc}$. Up until about 3 kHz (limited by the low pass filters along the DC bias circuit), $\Gamma^{(dark)}_{meas}$ increases with $I_b$ due to a stronger escape tendency of the phase particle as the barrier height $\Delta U \simeq (4\sqrt{2}/3)(\hbar I_c/2e)(1-I_b/I_c)^{3/2}$ decreases, where $I_c$ is the JJ critical current, $\hbar$ and $e$ are the reduced Planck constant and electron charge, respectively. The nearly linear dependence of $\Gamma^{(dark)}_{meas}$ in the semilog plot agrees with an escape rate $\Gamma^{(dark)}_{esc} \propto e^{-\Delta U/k_BT_{esc}}$ with $k_B$ and $T_{esc}$ being the Boltzmann constant and escape temperature, respectively. This is a characteristic exponential form in reaction-rate theory with an activation energy of $k_BT_{esc}$ \cite{Mannik:2005fua}. For TA escape, $k_BT_{esc}$ is given by the thermal energy whereas for MQT escape, by the energy of the harmonic oscillator at the local minima of the tilted-washboard potential, i.e. $k_BT_{esc}\simeq \hbar\omega_P/7.2 (1+0.87/Q)$ with $\omega_P$ and $Q$ being the JJ plasma frequency under current bias and the quality factor of the harmonic potential, respectively. $\Gamma^{(dark)}_{meas}$ is best fit by the MQT theory (black solid line) with $I_c$ of 11.99 $\mu$A such that $\omega_{P}/2\pi = 225$ GHz at zero $I_b$ and $Q\simeq 1.1$ (SI).

While $\Gamma^{(dark)}_{meas}$ = $\Gamma^{(dark)}_{esc}$, $\Gamma^{(light)}_{meas}$ in the light does not directly measure $\Gamma^{(light)}_{esc}$, i.e. the single-photon-enhanced escape rate. For a brief moment $\Delta t_1$ when $\Gamma^{(light)}_{esc} \gg \Gamma^{(dark)}_{esc}$, the phase particle escape probability is $\mathcal{P}_{esc}^{(light)} = 1-\exp(-\int_0^{\Delta t_1}\Gamma^{(light)}_{esc} dt)$. To simplify our calculation, we consider a constant $\Gamma^{(light)}_{esc}$ during $\Delta t_1$ and focus on the lower $I_b$ regime where $\mathcal{P}_{esc}^{(light)} \simeq \Gamma^{(light)}_{esc} \Delta t_1 \ll 1$. With negligible $\Gamma_{meas}^{(dark)}$, $\Gamma^{(light)}_{meas} \simeq \mathcal{P}_{esc}^{(light)}\mathcal{R}_{photon} \simeq \Gamma^{(light)}_{esc} \Delta t_1 \mathcal{R}_{photon}$. Assuming $\Delta t_1$ does not depend on $I_b$ exponentially, the nearly linear $\Gamma_{meas}^{(light)}$ in the semilog plot suggests $\Gamma_{esc}^{(light)}$ follows the reaction-rate theory as the single-photon induced JJ switching can be described by an enhanced activation energy, $k_BT^*_{esc}$. More quantitatively, we fit the data using $T^*_{esc}$ and $\Delta t_1$ as two free parameters accounting for MQT, TA, and phase diffusion processes \cite{Mannik:2005fua} (SI). Color solid lines in Fig. 3B show the results with $T^*_{esc} \simeq 2.1$ K and $\Delta t_1 \simeq 0.86$ ns. We attribute $T^*_{esc}$ to an elevated effective temperature ($>T_{esc}$) caused by QPs generated in the superconducting contacts that will be discussed later. The close fit of the model to the data indicates a photon induces JJ switching by momentarily elevating the activation energy.

\begin{figure*}
\includegraphics[width=7in]{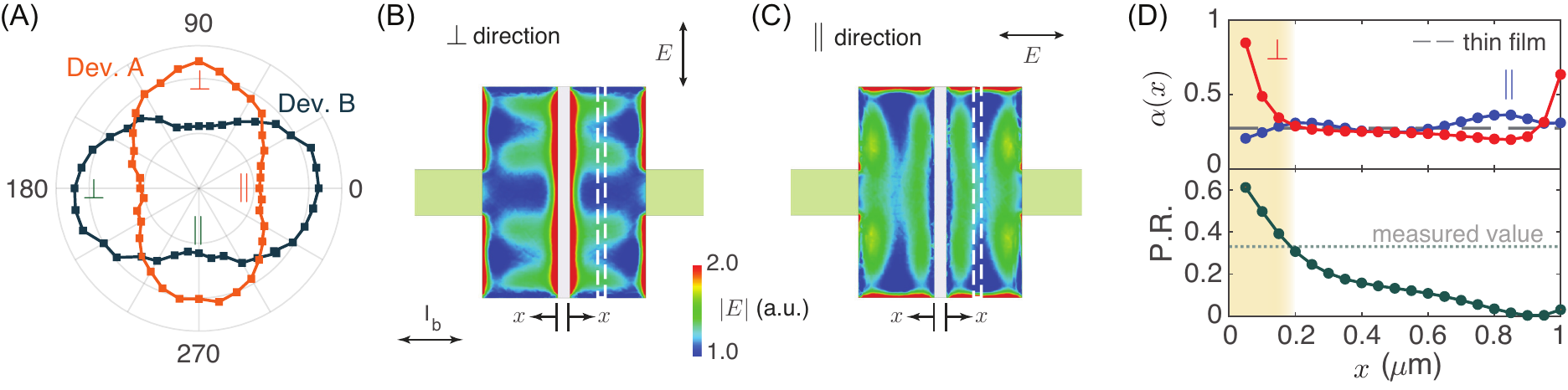}
\caption{\textbf{NIR photon coupling into Josephson junction. (A)} The polarization dependence of $\Gamma_{meas}$ (radial) for Device A and B (same orientation as the optical image in Fig. 1B) are orthogonal to each other, demonstrating that the single-photon switching is due to photons absorbed directly at the JJs. \textbf{(B-C)} The simulated electric field strength for incident light perpendicular and parallel to the supercurrent direction, respectively. The absorption occurs in the NbN contacts, enhanced through coupling to localized surface plasmons (red). \textbf{(D)} Upper panel: averaged volumetric NIR absorption $\alpha$ of the two polarizations versus distance $x$ from the interface between the graphene and superconductor. Dashed line is the value of $\alpha$ for thin film NbN. Lower panel: the polarization ratio $[\int^x_0(\alpha_\perp-\alpha_\parallel)dx\prime]/[\int^x_0(\alpha_\perp+\alpha_\parallel)dx\prime]$ versus $x$. The dotted line marks the measured ratio 33\% from (A), suggesting an effective absorption length of 190 nm (yellow shaded region) in NbN.}
\end{figure*}
To understand how the JJ absorbs single photons, we study the dependence of $\Gamma^{light}_{meas}$ on the polarization of incident light \cite{Driessen:2009kq}. Fig. 4A plots this angular dependence for two equally-sized JJs oriented orthogonally to one another on the same substrate (Fig. 1B) with arbitrarily oriented superconducting electrodes connecting the JJs to wire-bonding pads. Both devices exhibited the same photon rate but offset by 90$^{\circ}$ to each other. This dependence on junction geometry suggests that the detected single photons are absorbed at the JJs; not the electrodes. By inserting and rotating a polarizer in front of the JJs in a series of experiments (SI), we find the maximum $\Gamma^{light}_{meas}$ coincides with the polarization perpendicular ($\perp$) to the JJ supercurrent flow.
 
Using ANSYS HFSS, we study the angular dependence by computing the volumetric photo-absorption for the entire junction area with an incident plane wave approximation because the JJ is much smaller than and located at the center of the Gaussian beam waist of 2.41 mm. In Fig. 4B and C, we plot the field distribution within the NbN. When the real part of the permittivity of NbN ($-6.52+60.88i$) \cite{Semenov:2009em} and its adjacent medium have opposite signs, they form effectively a capacitor-inductor network that supports a localized surface plasmon along the interface \cite{Engheta:2007ca}. With a photon polarization in $\perp$ direction, the plasmon coupling intensifies the field and enhances the photon absorption at the graphene-NbN interface, resulting in the observed polarization-dependent $\Gamma_{meas}$.

We can estimate the effective area of the JJ as a SPD by matching the measured and simulated polarization ratios. By defining volumetric absorption zones (dashed line in Fig. 4B and C) at distance $x$ from the graphene-NbN interface with a 50 nm $\times 2.8 \mu$m area, Fig. 4D plots the spatial dependence of photon absorption coefficient $\alpha_j(x)$, where $j = \perp, \parallel$ notates the polarization direction. $\alpha_\perp$ and $\alpha_\parallel$ differ considerably near the edge due to the surface plasmon, but approach the same value, $\alpha = 0.27$ (dashed line), further away from the edge. This value agrees with the expected photo-absorption of 50 nm thin film NbN \cite{Driessen:2009kq} and verifies our calculation. Photons could be absorbed directly into the graphene at a rate of 0.07 Hz and 0.27 Hz with 100 pW of laser power in the $\perp$ and $\parallel$ directions, respectively. This is much smaller than $\Gamma^{light}_{meas}$ and orthogonal to the measured polarization dependence, thus excluding the detected single photons being absorbed directly in graphene. We define the polarization ratio as $[\int^x_0(\alpha_\perp-\alpha_\parallel)dx\prime]/[\int^x_0(\alpha_\perp+\alpha_\parallel)dx\prime]$ \cite{Driessen:2009kq}, that sums up the photon absorption with increasing distance from the graphene-NbN interface. To match the measured polarization ratio of 33\% from Fig. 4A, the simulation suggests the photons absorbed within 190 nm of the edge are those that trigger the JJ switching events. Therefore the total effective area of our JJ SPD is about 1.06 $\mu$m$^2$ with an averaged $\alpha$ of 0.58 in the $\perp$ polarization direction, resulting in the $\mathcal{R}_{photon}$ value (SI) used for calculating $\langle N_{photon}\rangle$ in Fig. 2B and fitting data in Fig. 3B.

Analogous to the QP-induced qubit relaxation \cite{Riste:2013il,Wang:2014dv,Serniak:2018dp}, we argue that QPs mediate the photon-induced JJ switching. When a single NIR photon impinges onto the superconductor, it breaks Cooper pairs and generates $\sim$261 QPs, given by $\eta \hbar\omega_{photon} / \Delta$ with $\omega_{photon}/2\pi$ being the photon frequency, $\Delta$ being the superconducting gap energy, and $\eta = 0.57$ being the downconversion efficiency \cite{Peacock:1996io}. Before recombining through inelastic scattering, these QPs can diffuse across the JJ, resulting in a diffusion current $I_{diff} = (e/4\sqrt{\pi\mathcal{D}t})xe^{(-x^2/4\mathcal{D}t)}$ per QP at the JJ with $\mathcal{D}$ = 0.55 cm$^2$/s being the diffusion constant in NbN \cite{Semenov:2009em}. For $x = 100$ nm, $I_{diff}$ is 11.5 nA which rises and subsides in a characteristic time scale of $x^2/\mathcal{D} \simeq 0.2$ ns. $I_{diff}$ is much smaller than the required $I_c - I_b$ to directly switch the JJ. However, unless scattered inelastically and trapped inside the JJ, these QPs at the gap energy can diffuse across the JJ with a mean-free-path of $\sim$92 nm in graphene through the 160-nm long channel. The JJ in our experiment is quasi-ballistic such that the $I_{diff}$ could generate a shot noise $S_I = 2e\mathcal{F}I_{diff}$ with a Fano factor $\mathcal{F}$ on the order of 1. Similar to how the current noise relaxes the current-biased JJ qubit by coupling through its shunt resistance \cite{Martinis:2003bq}, shot noise can enhance JJ switching by exciting the phase particle to higher energy states thus increasing its escape probability \cite{Pekola:2005iu}. This mechanism is equivalent to the reaction-rate theory with an effective temperature $T^*$ given by \cite{Pekola:2005iu} $\hbar\omega_P/k_B[2\coth^{-1}{(1+QS_I(\omega_P)/\hbar\omega_P^2C_{JJ})}]$ with $C_{JJ} = 20$ fF being the effective JJ capacitance \cite{Angers:2008gm}. For $\mathcal{F}=1$, $T^*\simeq$ 1 K, the same order of magnitude but a factor of two lower than our fitted value of $T^*_{esc}$ in Fig. 3B. We note that $I_{diff}$ might rise above 1 $\mu$A for small $x$, i.e. 10 nm. However, the characteristic time scale of $I_{diff}$ is much faster than $1/\omega_P$, invalidating the use of the $T^*$ equation \cite{Pekola:2005iu}. In the future, theoretical calculations of $\Gamma^{light}_{esc}$ due to both adiabatic and non-adiabatic change of washboard potential will inform the microscopic mechanism of single-photon induced switchings and optimize the SPD efficiency. The role of graphene is inessential under this hypothetical mechanism except that it provides a shunt resistor, quasi-ballistic channel across the JJ and forms a proximity JJ that allows for the efficient coupling of photons by the dissipative surface plasmon. This is consistent with the lack of large dependence of gate voltage and graphene thickness in subsequent experiments (SI). Since we can control the resistance of our graphene-based JJ SPD, this feature will allow the matching of load impedance with JJ-based computer architectures to enable high-speed, low-power JJ optical interconnects.

\textbf{Acknowledgements.} We thank valuable discussions with L.~Levitov, M.~Shaw, T.~Heikkil\"{a}, and L.~Govia. The work of E.D.W. and D. E. was supported in part by the Army Research Laboratory Institute for Soldier Nanotechnologies program W911NF-18-2-0048 and the US Army Research Laboratory (Award W911NF-17-1-0435). W.J. and G.-H.L. were supported by National Research Foundation of Korea (NRF) funded by the Korean Government (grant no. 2016R1A5A1008184, 2020R1C1C1013241, 2020M3H3A1100839), Samsung Science and Technology Foundation (project no. SSTF-BA1702-05) and Samsung Electronics Co., Ltd. D.K.E. acknowledges support from the Ministry of Economy and Competitiveness of Spain through the ``Severo Ochoa'' program for Centres of Excellence in R\&D (SE5-0522), Fundaci\'{o} Privada Cellex, Fundaci\'{o} Privada Mir-Puig, the Generalitat de Catalunya through the CERCA program, the H2020 Programme under grant agreement 820378, Project: 2D$\cdot$SIPC and the La Caixa Foundation. B.-I. W. is based upon work supported by the Air Force Office of Scientific Research under award number FA9550-16RYCOR290. K.W. and T.T. acknowledge support from the Elemental Strategy Initiative conducted by the MEXT, Japan ,Grant Number JPMXP0112101001,  JSPS KAKENHI Grant Number JP20H00354 and the CREST(JPMJCR15F3), JST. P.K. and K.C.F. were supported in part by Army Research Office under Cooperative Agreement Number W911NF-17-1-0574.

\section{Supplementary Information}
\renewcommand{\thetable}{S\arabic{table}} 
\renewcommand\thefigure{S\arabic{figure}} 
\begin{table*}[t]
\centering
\begin{tabular}{ l c c c c c c c } 
\hline
Device &~~~~~~A~~~~~~&~~~~~~B~~~~~~&~~~~~~C~~~~~~&~~~~~~D~~~~~~&~~~~~~E~~~~~~&~~~~~~F~~~~~~&~~~~~~G~~~~~~\\ 
\hline
% Nickname & Woody JJ90 & Woody JJ2p8 & Woody JJ1p5 & Buzz & Stitch JJq & NemoJJ0 & NemoJJ1\\
%Chip serial number & BGB71 & BGB71 & BGB71 & BGB38 & BGB86 & BGB57 & BGB57\\ 
JJ width ($\mu$m) & 2.8 & 2.8 & 1.5 & 1.5 & 2.8 & 1.5 & 1.5\\ 
JJ channel length (nm) & 160 & 160 & 160 & 160 & 160 & 160 & 160 \\
Graphene layer & 1 & 1 & 1 & 1 & 4 & 1 & 1 \\
NbN stack (Ti/Nb/NbN in nm) & 5/2.5/50 & 5/2.5/50 & 5/2.5/50 & 5/5/100 & 5/5/75 & 5/5/50 & 5/5/50 \\
 $\langle I_s\rangle$ ($\mu$A)& 10.91 & 10.77 & 5.07 & 3.11 & 9.77 & 1.71 & 1.55 \\
 Normal resistance, $R_n$ ($\Omega$)& 44 & 27.3 & 70.8 & 62.5 & 27 & 101 & 94.1 \\
$\Delta V_{gate}$(V) & 26.5 & 26 & 25 & 24.5 & 20 & 39.6 & 35 \\
 $V_{CNP}$ (V) & -6.5 & -6 & -5 & -4.5 & n.a. & -9.6 & -5 \\
 Measured polarization ratio & 0.33 & 0.33 & 0.18 & 0.54 to 0.75 & 0.33 & n.a. & n.a. \\
\hline
\end{tabular}
\caption{\textbf{List of measured devices.}
The graphene-based JJs are fabricated using dry-transfer technique with one-dimensional edge contact \cite{Wang:2013ch,CastellanosGomez:2014bh,Calado:2015fp,Walsh:2017kk,Lee:2020ci}. The JJ channel lengths measured by scanning electron microscope are typically 40 nm shorter than the design value of 200 nm in lithography. $V_{CNP}$ is the gate voltage of the charge neutrality point for the monolayer graphene. $\Delta V_{gate}$ is the gate voltage, $V_{gate}$, measured from $V_{CNP}$}
\label{tab:Devices}
\end{table*}

\begin{table*}
\centering
\begin{tabular}{ l c c c c } 
\hline
Device & A & B & C & D \\
\hline
%Nickname & Woody JJ90 & Woody JJ2p8 & Woody JJ1p5 & Buzz \\
Electron density (10$^{12}$ cm$^{-2}$) & 1.99 & 1.98 & 1.89 & 1.67 \\
Electronic mobility (10$^{12}$ cm$^{2}$/V s) & 5588 & 7740 & 7739 & 8000 \\
Mean free path (nm) & ~~~~~~91.7~~~~~~ & ~~~~~~121~~~~~~ & ~~~~~~124~~~~~~ & ~~~~~~120~~~~~~ \\
$I_c$ ($\mu$A) & 11.99 & 11.47 & 3.78 & 3.54 \\
$R_n$ ($\Omega$) & 44 & 40 & 71 & 63 \\
$I_cR_n$ ($\mu e$V) & 528 & 459 & 269 & 223 \\
Thouless energy (m$e$V) & 0.76 & 0.74 & 1.6 & 0.99 \\
JJ coupling energy (m$e$V) & 25 & 24 & 7.8 & 7.25 \\
$\omega_{P0}/2\pi$ (GHz)& 225 & 200 & 500 & 156 \\
$C_{JJ}$ (fF)& 18 & 22 & 5.8 & 11 \\
$Q_0$ & 1.12 & 1.22 & 1.67 & 0.46 \\
$\Delta$ of NbN (m$e$V) & \multicolumn{4}{c}{1.52}\\
\hline
\end{tabular}
\caption{\textbf{List of JJ properties.} Values taken at gate voltage of 20 V.}
\label{tab:JJParameters}
\end{table*}

\subsection{Calculation of $\mathcal{R}_{photon}$ from laser power}\label{secRphoton}
To understand the measured single-photon induced JJ switching rate quantitatively, we calculate the number of incident photons per unit time per unit area per unit incident laser power through the optical fiber, i.e. $\mathcal{J}_{photon}$. We can calculate $\mathcal{J}_{photon}$ using the Gaussian beam profile as follows: The light that illuminates the JJ is guided by a single-mode fiber (SMF-28 designed for 1550-nm transmission). The radius of the laser beam, $w$, a distance $z$ away from the end of the fiber, assuming Gaussian beam propagation, is:
\ba w(z) = w_0\sqrt{1+(z/z_R)^2} \ea
where $w_0 = 5.2$ $\mu$m is the beam radius at the end of the fiber and  $z_R = \pi w_0^2/\lambda = 55$ $\mu$m is the Rayleigh range with $\lambda$ being the wavelength, i.e. 1550~nm. The device is 1 inch away from the end of the fiber resulting in $w$ = 2.41 mm at the JJ with a spot size of 18.25 mm$^2$. For a Gaussian beam, the intensity profile, $I$, as a function of $z$ and distance $r$ from the beam center is given by:
\ba I(r, z) = I_0\left(\frac{w_0}{w(z)}\right)^2 e^{-2(r/w(z))^2}\ea, where $I_0 = 2P_{laser}/(\pi w_0^2)$ with $P_{laser}$ being the total power of the beam given by the normalization condition, i.e. $\int_0^\infty I(r) 2\pi r dr = P_{laser}$. Therefore, at the center of the beam, $I(r=0, z) = 2P_{laser}/\pi [w(z)]^2$. Assuming the JJ is centered in the laser spot, 100 pW out of the fiber transduces to 10.96 pW/mm$^2$ at the JJ. Using 0.80 eV for 1550-nm photons, we calculate $\mathcal{J}_{photon}$ = 85.5 photons per second per $\mu m^2$ per 100 pW.

Base on this $\mathcal{J}_{photon}$, the effective single-photon absorption area, $A_{eff}$, estimated from the polarization ratio measurement and HFSS simulation, and averaged photon absorption coefficient, $\langle\alpha\rangle$, the expected absorbed photon rate, $\mathcal{R}_{photon}$, used in the main text, which is given by
\ba \mathcal{R}_{photon} = \mathcal{J}_{photon} A_{eff} \cdot \langle\alpha\rangle \cdot P_{laser}\ea
, are shown in Table \ref{tab:SPCounts}.
\begin{table*}
\centering
\begin{tabular}{ p{7.4cm}  c  c  c  c  } 
\hline
Device & A &  B & C & D\\ 
\hline
Measured polarization ratio & 0.33 & 0.33 & 0.18 & $0.65\pm 0.15$\\ 
Effective single-photon absorption area, $A_{eff}$, ($\mu$m$^2$) & ~~~$2\times 2.8 \times 0.19$~~ & ~$2\times 2.8 \times 0.19$~~ & ~$2\times 1.5 \times 0.27$~~ & ~$<2\times 1.5 \times 0.05$~~\\ 
Averaged coupling efficiency within $A_{eff}$, $\langle\alpha\rangle$ & 60\% & 60\% & 52\% & 94\% \\
$\mathcal{R}_{photon}$ with 100 pW laser power & 53 Hz & 53 Hz & 36 Hz & $<$ 12 Hz\\
\hline
\end{tabular}
\caption{Estimated photon rate, $\mathcal{R}_{photon}$, based on the polarization ratio measurement and HFSS simulation}
\label{tab:SPCounts}
\end{table*}

\subsection{Single-photon detection in pulsed measurements}
Using a pulsed laser as the photon source, instead of a continuous wave (CW) laser, can provide an independent method to cross-check the single-photon response of the JJ by: 1) showing that the detector responds to pulses with less than one photon on average and 2) verifying the linearity of the single-photon sensitivity with incident laser power for various pulse conditions \cite{Goltsman:2001eaa}. In the pulse measurement, we can control the number of photons reaching the JJ per pulse by the pulse laser power, $P_{pulse}$, and the pulse duration, $t_{pulse}$. The number of photons per pulse, $N_{pulse}$ is given by $N_{pulse}=E_{pulse}/E_{ph}=P_{pulse}t_{pulse}/E_{ph}$ where $E_{pulse}$ is the energy per pulse and $E_{ph}$ is the photon energy. By tuning $P_{pulse}$ and $t_{pulse}$ we can keep $N_{pulse}<1$ to ensure the JJ is sensitive to single photons. Furthermore, if the JJ is acting as a single-photon detector then its switching rate must be linear with the incident photon rate and therefore power. For a pulsed measurement, if this is true then as we tune $P_{pulse}$ we should be able to keep the probability of switching per pulse constant by inversely proportionally tuning $t_{pulse}$.

\begin{figure}[h]
\includegraphics[width=0.7\columnwidth]{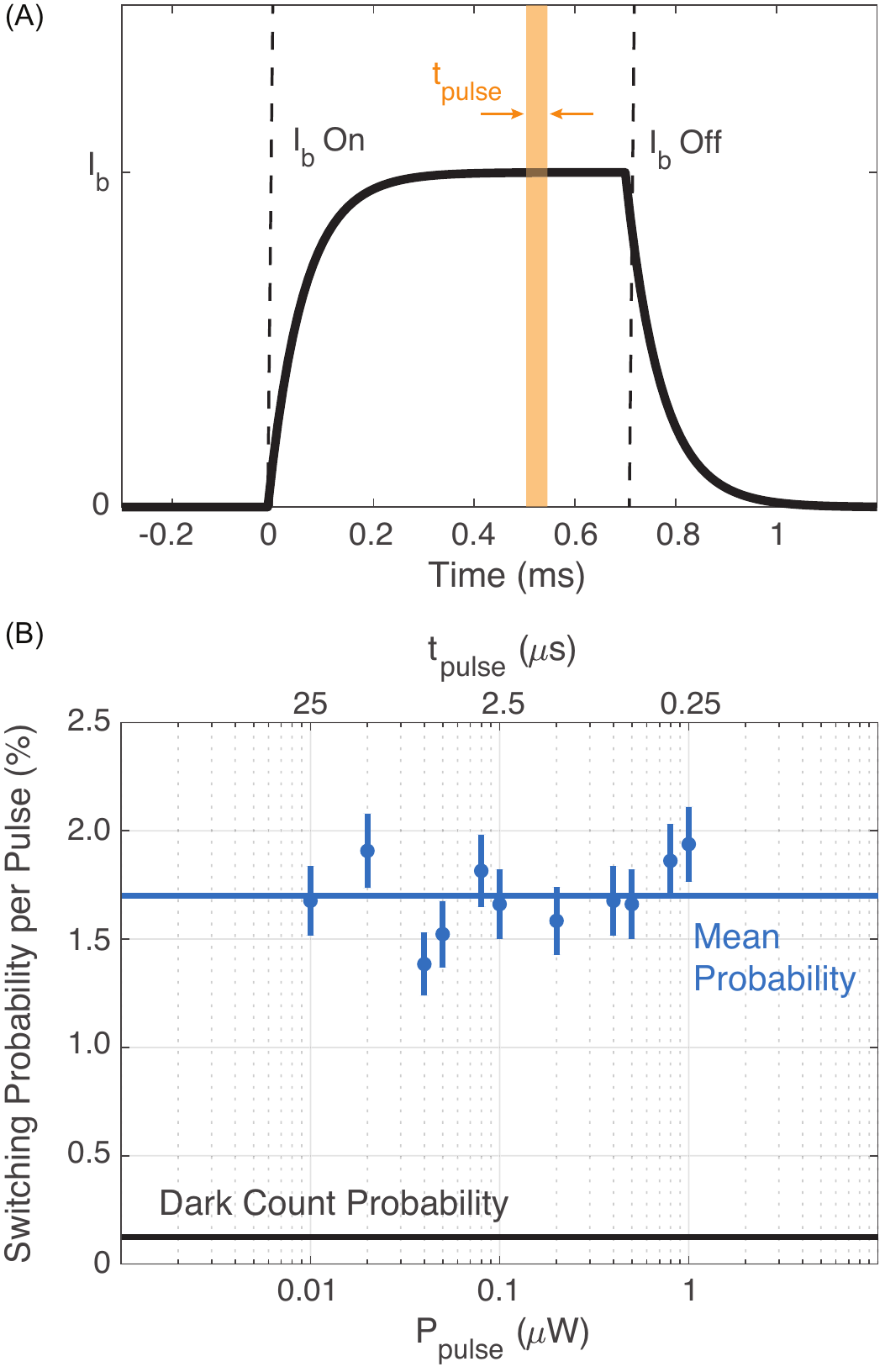}
\caption{\textbf{Single-photon-induced switching in a pulsed laser experiment} (A) The timing of the pulsed experiment. The laser pulse starts 500 $\mu$s after the bias current is turned on to allow for a $\sim$300 $\mu$s ramp-up time. The bias current is turned off 200 $\mu$s after the laser pulse starts to allow for any photon-induced switching events to be measured. The bias current is 10.70 $\mu$A for the data of this figure. The pulse repetition rate was 1 Hz so that we can completely eliminate any switching events caused by Joule heating. (B) Switching probability per pulse of Device A. If the energy per pulse, or equivalently the number of photons per pulse, is kept constant (in this case 250 fJ per pulse), the measured probability of switching per pulse remains constant for various powers (bottom axis) and their corresponding pulse duration (top axis). The average switching probability across trials is 1.7\% per pulse (blue line). The error bars are calculated as the square root of the measured number of counts divided by the number of pulses.}
\label{fig:pulseMeasurement}
\end{figure}

For the pulse experiment, we illuminate the JJ at a pulse repetition rate of 1 Hz. Such a low repetition rate can guarantee a sufficient cooling time between pulses to eliminate any Joule heating effect from frequent switching. Additionally, we bias the junction for a short time, i.e. 700 $\mu$s, in each pulse cycle to reduce the time window for dark counts. As shown schematically in Fig. \ref{fig:pulseMeasurement}A, we turn the bias current on 500 $\mu$s before the laser pulse to allow for a 300 $\mu$s turn on time (limited by the low pass filters in the setup), and then turn it off 200 $\mu$s after the laser pulse. For these measurements, $I_b$=10.70~$\mu$A. The voltage across the JJ is recorded to measure the probability of single-photon induced switching.

Measured on Device A using 6500 pulses for each data point, Fig. \ref{fig:pulseMeasurement}B plots the switching probability per pulse for eleven $(P_{pulse}, t_{pulse})$ pairs such that $E_{pulse}$ is held constant at 250 fJ. Using the result from Section \ref{secRphoton}, the number of photons per pulse reaching the JJ is:
\ba N_{pulse} &=& \mathcal{J}_{photon} A_{eff}\cdot\langle\alpha\rangle\cdot P_{pulse} t_{pulse}\nonumber\\
&=& (53 \text{Hz}/100 \text{pW})(250 \text{fJ})\nonumber\\ &=& 0.13\ea Assuming a Poisson distribution, the probability of zero, one, and two photons in each pulse are 0.878, 0.114, and 0.007, respectively. Therefore, most of the observed light-induced switching events are caused by single photons. The average switching probability across trials is 1.7\% per pulse. The error bars are given by the square root of the number of counts (the standard deviation assuming a Poisson distribution) divided by the number of pulses. The result shows a constant switching rate over two-orders of magnitude in the pulse power up to 1 $\mu$W, i.e. $10^4$ times larger than the CW power used in the main text.

\begin{figure}[h]
\includegraphics[width=1\columnwidth]{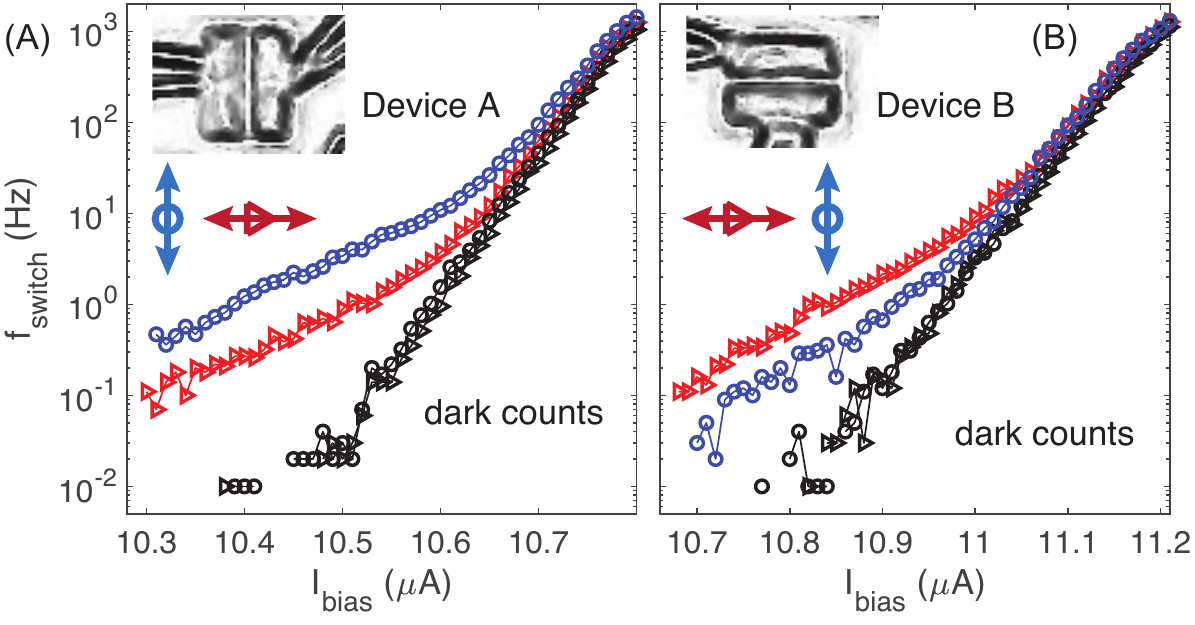}
\caption{\textbf{Determination of polarization orientation.} Measurements of the switching rate of two orthogonally-oriented JJs (Devices A and B in (A) and (B), respectively) are carried out with a linear polarizer between the light source and the devices. The experiment is repeated twice with the orientation of the polarizer rotated 90$^\circ$ between the two. The triangular data corresponds to one experiment and the circular data to the second. The black data is the dark count rate. The red and blue data is the switching rate in the presence of light. The double arrows show the orientation of the incident polarization relative to the JJ depicted in the inset. We observe for both devices that the switching rate is enhanced when the polarization is aligned perpendicular to the supercurrent direction (blue in (A) and red in (B)).}
\label{fig:determine_polarization_orientation}
\end{figure}

\subsection{Experimental determination of polarization orientation}
We perform a series of experiments to determine the polarization orientation of our incident photon relative to the orientation of the JJ. We do not use a polarization-maintaining fiber in our setup so the linear polarization of the light out of the fiber, and thus incident on the JJ, is not known a priori. To determine the polarization orientations causing the highest and lowest $\Gamma^{(light)}_{meas}$ (Fig. 4A in the main text), we add a linear polarizer between the output of the single-mode fiber and the JJs. The orientation of the JJs compared to their packaging is known and used to align the polarizer either parallel or perpendicular to the direction of the supercurrent flow in two separate cool-downs of the devices. To align the polarization orientation of the incoming light with the polarizer, we rotate the polarization while simultaneously measuring the switching rate of the JJs. The polarization is rotated until the switching rate of the JJs is at a maximum for a given laser intensity, corresponding to the maximum amount of light reaching the JJs and therefore the maximum alignment with the polarizer. At this polarization, the switching rate is measured as a function of bias current for both JJs. The results are displayed in Fig. \ref{fig:determine_polarization_orientation}  where the blue data is taken in one cool-down for both devices and the red in another. We observe that for both devices, the switching rate is highest when the light is polarized perpendicular to the supercurrent direction which is in agreement with our HFSS model.

\subsection{HFSS simulation}
To understand the polarization ratio and the photon coupling to the JJ, we model the device using finite element analysis in ANSYS HFSS. The model is based on the geometry of Devices A, B, C, and D. The dimensions, thickness, and materials are listed in Table 1. The NbN contacts have the same width as the graphene and length 1 $\mu$m. The hBN-encapsulated graphene and NbN/Nb/Ti stack are on 285 nm of SiO$_2$ with 500 $\mu$m of p-doped Si below that. To limit the boundary effects, periodic boundary conditions are used on the four surrounding sides and an impedance boundary is applied at the bottom side to terminate the silicon layer. Regarding the constitutive parameters of the materials at 1.55 $\mu$m, we use: $\epsilon_{NbN} = -6.52+60.88i$ \cite{Kerman:2008cr, Driessen:2009kq}, $\epsilon_{Nb} = -72.39+7.67i$, and $\epsilon_{Ti} = -7.67+33.92i$ for the key lossy structures above the silicon dioxide. Normally incident plane wave with polarization angles of 0$^{\circ}$ and 90$^{\circ}$ , corresponding to the electric field of photons in parallel and perpendicular to the supercurrent flow respectively, are used. The volumetric absorption of the NbN/Nb/Ti stacks can be calculated numerically within the simulation. We shift the distance of the volumetric absorption zone from the Josephson junction in the direction of supercurrent to obtain the data in Fig. 4D in the main text. The white dashed lines in Fig. 4B and C in the main text outline the volumetric absorption zone which has a dimension of $0.05 \mu m \times 2.8 \mu m$ for Device A and B, and of $0.05 \mu m \times 1.5 \mu m$ for Device C and D. 

\begin{figure}[h]
\includegraphics[width=0.65\columnwidth]{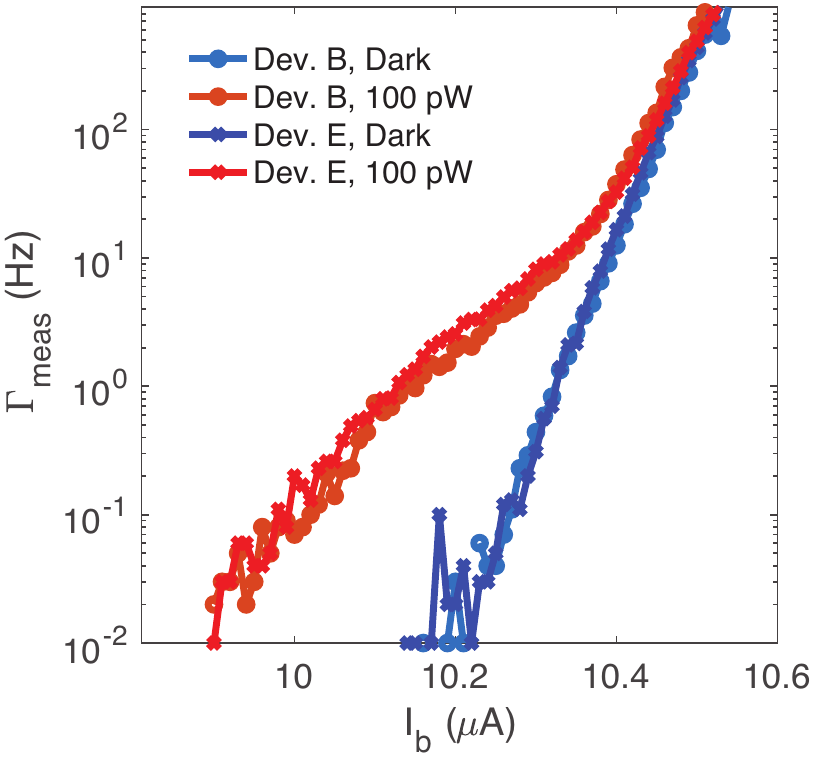}
\caption{\textbf{$\Gamma_{meas}$ of JJ using tetra-layer graphene.} Switching rate of a JJ with tetra-layer graphene (Device E) compared to the one with monolayer graphene (Device B) at $V_{gate}$ = 20 V and $T$ = 36 mK. The switching rate is nearly identical between the two devices pointing to the the fact the absorption takes place in the NbN contacts instead of in the graphene itself and the mechanism of single-photon induced switching does not depend critically on the graphene thickness. Device E data is offset by 0.65 $\mu$A to account for its difference of $I_c$ when compared with Device B.}
\label{fig:4layer}
\end{figure}

\subsection{$\Gamma_{meas}$ of JJ using tetra-layer graphene}
To further investigate the role of graphene in the photon absorption and observed JJ switchings, we fabricate a JJ with tetra-layer graphene (Device E) with dimensions comparable to Device A and B. The measurement results are plotted in Fig. \ref{fig:4layer} with an offset of 650 nA in $I_b$ for Device E to account for the difference in critical current from Device A. $\Gamma_{meas}$ for the two devices is nearly identical, in both the single-photon counts as well as its dependence on $I_b$. This result is consistent with the HFSS simulation that the photon absorption is dominated by the localized surface plasmon at the superconducting electrode right at the JJ. It also suggests that the mechanism of the single-photon induced switching does not depend critically on the graphene thickness.

\begin{figure}[h]
\includegraphics[width=0.65\columnwidth]{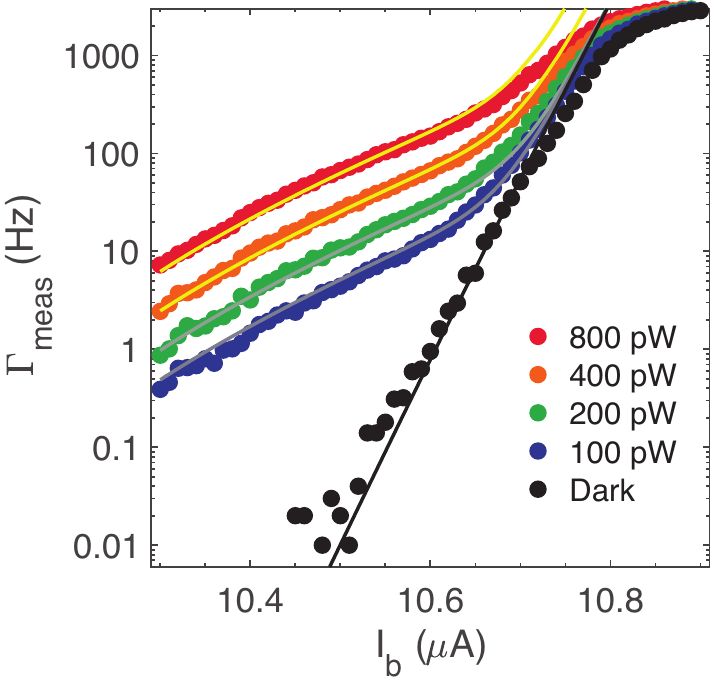}
\caption{\textbf{$\Gamma_{meas}$ of Device B} The switching rate as a function of bias current in dark and for various incident laser powers for Device B, similar to Fig. 3B in the main text. The fitting parameters for Device B were $T^*_{esc}$=1.80 K and $\Delta t_1$=0.668 ns, similar to the values for Device A.}
\label{fig:3B_DeviceB}
\end{figure}

\subsection{Modeling of $\Gamma_{meas}$ in Fig. 3B}\label{secModel}
In the main text, it is noted that the switching rate of the JJ is modeled using a phase particle model with escape rate given by \cite{Devoret:1985jx}:
$\Gamma_{esc} = Ae^{-\Delta U/k_BT_{esc}}$, where
\ba
A=\begin{cases}
\frac{\omega_p}{2\pi}\left(\sqrt{1+\frac{1}{4Q^2}}-\frac{1}{2Q}\right) &\text{(TA)}\\
12\omega_p\sqrt{\frac{3\Delta U}{2\pi \hbar \omega_p}} &\text{(MQT)}
\end{cases}
\ea
\ba
T_{esc}=\begin{cases}
T &\text{(TA)}\\
\hbar \omega_p/\left[7.2k_B\left(1+\frac{0.87}{Q}\right)\right] &\text{(MQT)}
\end{cases}
\ea
with $\omega_p=\omega_{p0}(1-\gamma_{JJ}^2)^{1/4}$ being the JJ plasma frequency, $\omega_{p0} = (2eI_c/(\hbar C_{JJ}))^{1/2}$ being the JJ plasma frequency at zero bias current, $I_c$ being the critical current, $\gamma_{JJ}=I_b/I_c$ being the normalized bias current, $C_{JJ}$ being the effective junction capacitance \cite{Angers:2008gm, Lee:2011et, Walsh:2017kk}, $Q=\omega_pR_nC_{JJ}$ being the JJ quality factor, $R_n$ being the normal-state resistance, $\Delta U = 2E_{J0}(\sqrt{1-\gamma_{JJ}^2}-\gamma_{JJ}\cos^{-1}{\gamma_{JJ}})$ being the phase-particle energy barrier, $E_{J0} = \hbar I_c/(2e)$ being the Josephson coupling energy, and $T$ being the JJ electron temperature. Here, $k_B$, $\hbar$, and $e$ are Boltzmann's constant, Planck's reduced constant, and the electron charge, respectively. To fit the data of count rate versus bias, we include both TA and MQT in in our model. We estimate an effective junction capacitance $C_{JJ}=$18 fF from the Thouless energy, $E_{Th}=\hbar D_e/L^2$, using $C_{JJ}=\hbar/R_nE_{Th}$. Here $D_e=v_Fl_{MFP}/2$ is the diffusion constant, $v_F\sim 10^6$ m/s is the Fermi velocity, and $l_{MFP}$ is the electron mean free path in graphene. We determine $I_c$ (which is in general $\sim$10\% higher than $\langle I_s\rangle$) by fitting the dark count rate with $\Gamma_{meas}^{(dark)}=\Gamma_{esc}^{(dark)}(I_c,T_0)=\Gamma_{MQT}(I_c)+\Gamma_{TA}(I_c,T_0)$ with $I_c$ as the only fitting parameter (here $T_0$=27 mK is the base temperature). For Device A, we find $I_c=$11.96 $\mu$A which determines the other junction parameters such as $E_{J0}$, $\Delta U$, $\omega_p$, and $Q$.

To model the enhanced switching probability induced by single photons, we assume the effective junction temperature remains at a constant enhanced-escape temperature $T^*_{esc}$ for some time $\Delta t_1$ upon photon absorption in the NbN contacts and subsequent quasiparticle diffusion. The enhanced-escape temperature is not a true temperature rise in the junction but instead is due to quasiparticle noise. This effective temperature will decay in time but we make the assumption that it is a constant during $\Delta t_1$ to simplify the model. The probability of the phase particle escaping during $\Delta t_1$ is $\mathcal{P}^{(light)}_{esc} = 1-\exp(-\int_0^{\Delta t_1}\Gamma^{(light)}_{esc} dt)\simeq 1-\exp(-\Gamma^{(light)}_{esc}\Delta t_1)$. For our range of operation we have $\Gamma^{(light)}_{esc}\Delta t_1\ll 1$ so $\mathcal{P}^{(light)}_{esc}\simeq\Gamma^{(light)}_{esc}\Delta t_1$. With $\mathcal{P}^{(light)}_{esc}$ being the probability of switching with each absorbed photon, the photon-induced switching rate in our setup is $\mathcal{P}^{(light)}_{esc}\mathcal{R}_{photon}\simeq\Gamma^{(light)}_{esc}\Delta t_1\mathcal{R}_{photon}$. Combining this with the dark count rate when no photons are present, we have for $\Gamma_{meas}^{(light)}$, i.e. the total measured switching rate of our setup under illumination: 
\begin{equation}\label{eqnGammaLight}
\begin{alignedat}{2}
\Gamma_{meas}^{(light)}&=&&\Gamma_{esc}^{(dark)}-\mathcal{R}_{photon}(1-e^{-\int_0^{\Delta t_1}\Gamma^{(dark)}_{esc} dt})\\
& &&+\mathcal{R}_{photon}(1-e^{-\int_0^{\Delta t_1}\Gamma^{(light)}_{esc} dt})\\
&\simeq&&\Gamma_{esc}^{(dark)}(I_c,T_0)(1-\Delta t_1\mathcal{R}_{photon})\\
& &&+\Gamma_{esc}^{(light)}(I_c,T^*_{esc})\Delta t_1\mathcal{R}_{photon}
\end{alignedat}
\end{equation}

We fit $\Gamma_{meas}^{(light)}$ versus $I_b$ for the two lowest light intensities that we study (100 and 200 pW which is $\sim$53 and $\sim$106 photons per second into the JJ) with $\Delta t_1$ and $T^*_{esc}$ as the fitting parameters and find good agreement with the data.

We can further improve the model by including the effect of phase particle retrapping events. As $I_b$ decreases, it gets closer to the retrapping current, $I_{r0}$, where the phase particle is retrapped and the JJ returns to the zero-voltage state. In practice, the phase particle will retrap at a current $I_r<I_{r0}$ due to thermal and quantum fluctuations (similar to the measured switching current $I_s$ being less than $I_c$). When retrapping becomes significant, the junction is said to be in the phase diffusion (PD) regime and shows reduced escape rate. Although there will still be a voltage spike in the case that the phase particle escapes and retraps, the spike will be transient so the JJ does not latch into the non-zero-voltage state, forbidding our setup from recording the event. The retrapping rate, $\Gamma_{r}$, can be calculated as \cite{BENJACOB:1982vo, Mannik:2005fua, Krasnov:2007eo, Bae:2009hy}:
\ba
\Gamma_r=\omega_{p0}\frac{I_b-I_{r0}}{I_c}\sqrt{\frac{E_{J0}}{2\pi k_BT}}e^{-\frac{\Delta U_r}{k_BT}}
\ea
with
\ba
\Delta U_r=\frac{E_{J0}Q^2}{2}\left(\frac{I_b-I_{r0}}{I_c}\right)^2
\ea
Now modifying Equation \ref{eqnGammaLight} and noting that $\Gamma_r^{(dark)}$ is negligible in our measurement range:
\begin{widetext}
\ba
\begin{split}
\Gamma_{meas}^{(light)}=&\Gamma_{esc}^{(dark)}(I_c,T_0)(1-\Delta t_1\mathcal{R}_{photon})\\
&+\Gamma_{esc}^{(light)}(I_c,T^*_{esc})\Delta t_1\mathcal{R}_{photon}(1-\Gamma_r^{(light)}(I_{r0},T^*_{esc})\Delta t_1)
\end{split}
\ea
\end{widetext}
We refit the data adding $I_{r0}$ as an additional fitting parameter to $\Delta t_1$ and $T^*_{esc}$. Here, we set an upper bound of $\Gamma_r\Delta t_1$ to be 1. The result is $\Delta t_1=$ 0.86 ns, $T^*_{esc}=$ 2.1 K, and $I_{r0}=$8.55 $\mu$A. For comparison, $\langle I_r\rangle =$6.64 $\mu$A, 78\% of  $I_{r0}$, showing reasonable agreement.  The model with these parameters agrees well with the data in Fig. 3B of the main text for all bias currents. We note that the model starts underestimating $\Gamma_{meas}^{(light)}$ at higher incident powers which we attribute to the heating of the substrate. We correct the fit by decreasing $I_c$ slightly at the higher incident powers. This heating correction to $I_c$ is small, requiring only a 0.4\% reduction in $I_c$ at the highest laser power that we used (800 pW).

Fig. \ref{fig:3B_DeviceB} shows the data and fitting of $\Gamma_{meas}$ for Device B using the same methods. We find $T^*_{esc}$=1.8 K and $\Delta t_1$=0.67 ns, similar to the values for Device A (see comparison in Table \ref{tab:3BFitParams}).
\begin{table}[h]
\centering
\begin{tabular}{ c c c} 
\hline
~~~Device~~~&~~~$T^*_{esc}$ (K)~~~&~~~$\Delta t_1$ (ns)~~~\\ 
\hline
A & 2.1 & 0.86\\
B & 1.8 & 0.67\\ 
\hline
\end{tabular}
\caption{\textbf{Fitting parameters for the single-photon response observed in Fig. 3B and Fig. S8.}}
\label{tab:3BFitParams}
\end{table}

\begin{figure*}[h]
\includegraphics[width=6.5in]{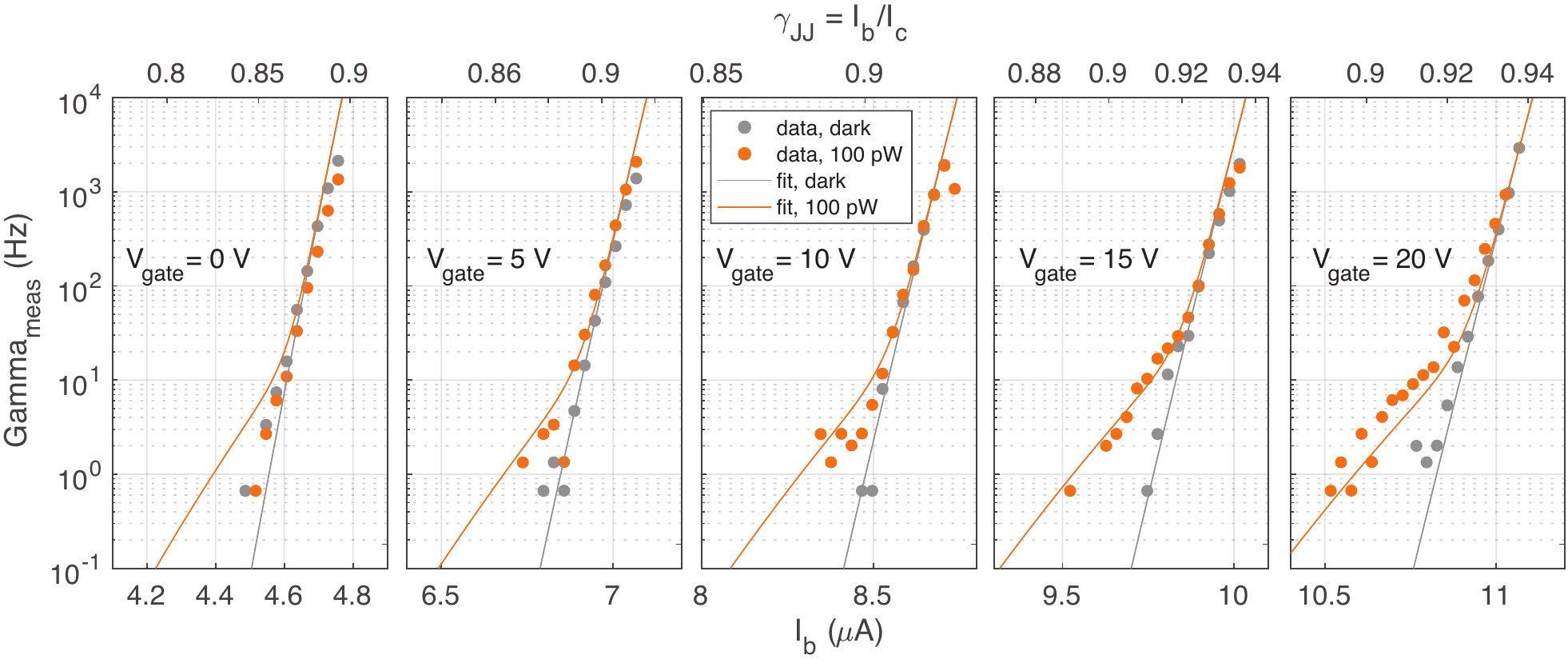}
\caption{\textbf{Gate-dependent switching rate of Device A} $\Gamma_{meas}^{(dark)}$ (gray circles) and $\Gamma_{meas}^{(light)}$ for 100 pW (orange circles) versus bias current $I_b$ (lower x-axis) and normalized bias current $\gamma_{JJ}$ (upper x-axis) for five different gate voltages. The gray and orange lines are the best fits for $\Gamma_{meas}^{(dark)}$ and $\Gamma_{meas}^{(light)}$, respectively, using the model of Section \ref{secModel}. For a given $\gamma_{JJ}$, $\Gamma_{meas}^{(light)}/\Gamma_{meas}^{(dark)}$ increases with increasing gate voltage so that the single-photon switching events can be observed more clearly at higher gate voltages. Data was collected using the bias-current sweep method \cite{Fulton:1974uf}.}
\label{fig:gateMQTvsTA}
\end{figure*}

\subsection{Gate Dependence of the JJ Single-Photon Response}
We study how the photon-induced switching depends on the gate voltage, $V_{gate}$. We measure the switching-current distribution of Device A for 5 different values of $V_{gate}$ (0 V to 20 V in steps of 5 V) using a 10 $\mu$A/s bias-current sweep both in the dark and with 100-pW of illumination and convert to $\Gamma_{meas}(I_b)$\cite{Fulton:1974uf}. As shown in Fig. \ref{fig:gateMQTvsTA} for Device A, the ratio $\tilde{\Gamma} = \Gamma_{meas}^{(light)}/\Gamma_{meas}^{(dark)}$ decreases as the gate voltage ($V_{gate}$) decreases for a given $\gamma_{JJ}=I_b/I_c$. This makes the observation of the single-photon induced switchings more favorable at high electron doping (i.e. $V_{gate}$) than at low doping. We attribute this behavior to two effects.

The first effect is the relative change in $\Gamma_{TA}$ compared to $\Gamma_{MQT}$ as a function of $V_{gate}$, with $\Gamma_{meas}^{(light)}$ dominated by TA and $\Gamma_{meas}^{(dark)}$ by MQT. In Fig. \ref{fig:gateMQTvsTA}, the y-axis limits are the same in all plots of $\Gamma_{meas}^{(light)}$ and $\Gamma_{meas}^{(dark)}$ at five different gate voltages. While the $I_c$ increases at higher $V_{gate}$, we also plot the normalized $I_b$ on the top x-axis as $\gamma_{JJ} = I_b/I_c$, which shifts to higher values as $V_{gate}$ increases. For a given value of $\gamma_{JJ}$, both $\Gamma_{MQT}$ and $\Gamma_{TA}$ decrease with increasing $V_{gate}$. However, $\Gamma_{MQT}$ decreases more quickly than does $\Gamma_{TA}$. Therefore, the MQT-dominated dark count rate falls faster than the TA-dominated single-photon response, making high gate voltage operation favorable. Qualitatively, $\Gamma_{MQT}$ is heavily determined by $\omega_p$ which depends on $I_c$, while $\Gamma_{TA}$ is determined more strongly by the temperature than $\omega_p$. Thus, the $\tilde{\Gamma}$ ratio is higher at higher gate voltages for a given $\gamma_{JJ}$. We fit the data using the same method as in Section \ref{secModel}, optimizing simultaneously over four different gate voltages: 5 V, 10 V, 15 V, and 20 V. $V_{gate}$ at the charge neutrality point, $V_{CNP}$, of the device is -6.5 V. The best fit gives $T^*_{esc}$=2.13 K and $\Delta t_1=$ 0.92 ns, in agreement with the fit from Fig. 3B. We plot these fits as lines for all five gate voltages.

Retrapping is the second cause of a lower $\tilde{\Gamma}$ ratio at lower gate voltage which can be observed in the $V_{gate}$=0 V case in Fig. \ref{fig:gateMQTvsTA}. Retrapping causes the observed count rate to be lower than it would otherwise be with only TA or MQT present. In Fig. \ref{fig:gateMQTvsTA}, at $V_{gate}$=0 V the data for $\Gamma_{meas}^{(light)}$ is essentially equal to the data for $\Gamma_{meas}^{(dark)}$. This result is in contrast to the best-fit line that suggests we should still have observed single-photon induced switching above the dark count rate. At lower gate voltages, the JJ enters the phase diffusion (PD) regime upon photon absorption so that retrapping becomes more significant and the count rate is reduced, similar to the reduced switching rate in the low-bias-current case of Section \ref{secModel}. 

%\bibliographystyle{apsrev}
%\bibliography{arXivV3}

\end{document}